# Evaluation of Compton scattering sequence reconstruction algorithms for a portable position sensitive radioactivity detector based on pixelated Cd(Zn)Te crystals

K. Karafasoulis[a,b], K. Zachariadou[c,d], C. Potiriadis[b], S. Seferlis[b], I. Kaissas[b], D. Loukas[d], C. Lambropoulos[e]

*a.Hellenic Army Academy, 16673 Vari, Greece*
*b.Greek Atomic Energy Commission, Patriarxou Grigoriou & Neapoleos Aghia Paraskevi Athens P.O Box 60092 Postal Code 15310*
*c.Hellenic Naval Academy, Xatzikiriakio, 18539 Piraeus, Greece*
*d. Institute of Nuclear Physics, National Center for Scientific Research Demokritos, P.O Box60228*
*e.Technological Educational Institute of Chalkida (TEI of Chalkida), Psachna Evias Postal Code 34400 Greece*

**Abstract.** We present extensive simulation studies on the performance of algorithms for the Compton sequence reconstruction used for the development of a portable spectroscopic instrument (COCAE), with the capability to localize and identify radioactive sources, by exploiting the Compton scattering imaging. Various Compton Sequence reconstruction algorithms have been compared using a large number of simulated events. These algorithms are based on Compton kinematics, as well as on statistical test criteria that exploit the redundant information of events having two or more photon interactions in the active detector's volume. The efficiency of the best performing technique is estimated for a wide range of incident gamma-ray photons emitted from point-like gamma sources.

## A. INTRODUCTION

COCAE is a portable spectroscopic system under development within the framework of the COCAE[1] project aimed to be used for the accurate localization and identification of radioactive sources and radioactively contaminated spots, in a broad energy range up to 2MeV.

The proposed portal instrument can be used for security inspections at the borders (airports, seaports etc) such as the determination of the position and the strength of potential radioactive sources. It can also be used at recycling factories for the detection of possible radioactive sources into scrap metals. Furthermore, the COCAE instrument could improve a lot the existing procedures at the nuclear waste management facilities as well as provide fast and accurate information during the response after a nuclear emergency situation.

[1] The project is funded under the European Community's FP7 -Program- Theme Security. Web Site http:// www.cocae.eu

For the source localization task, COCAE exploits the Compton scattering imaging, a technique widely used in many fields such as nuclear medicine, astrophysics and recently for counterterrorism. Among the main advantages of the proposed system is its enhanced efficiency and high energy resolution in combination with its independency of any cryogenic system.

COCAE is schematically shown in Figure 1.

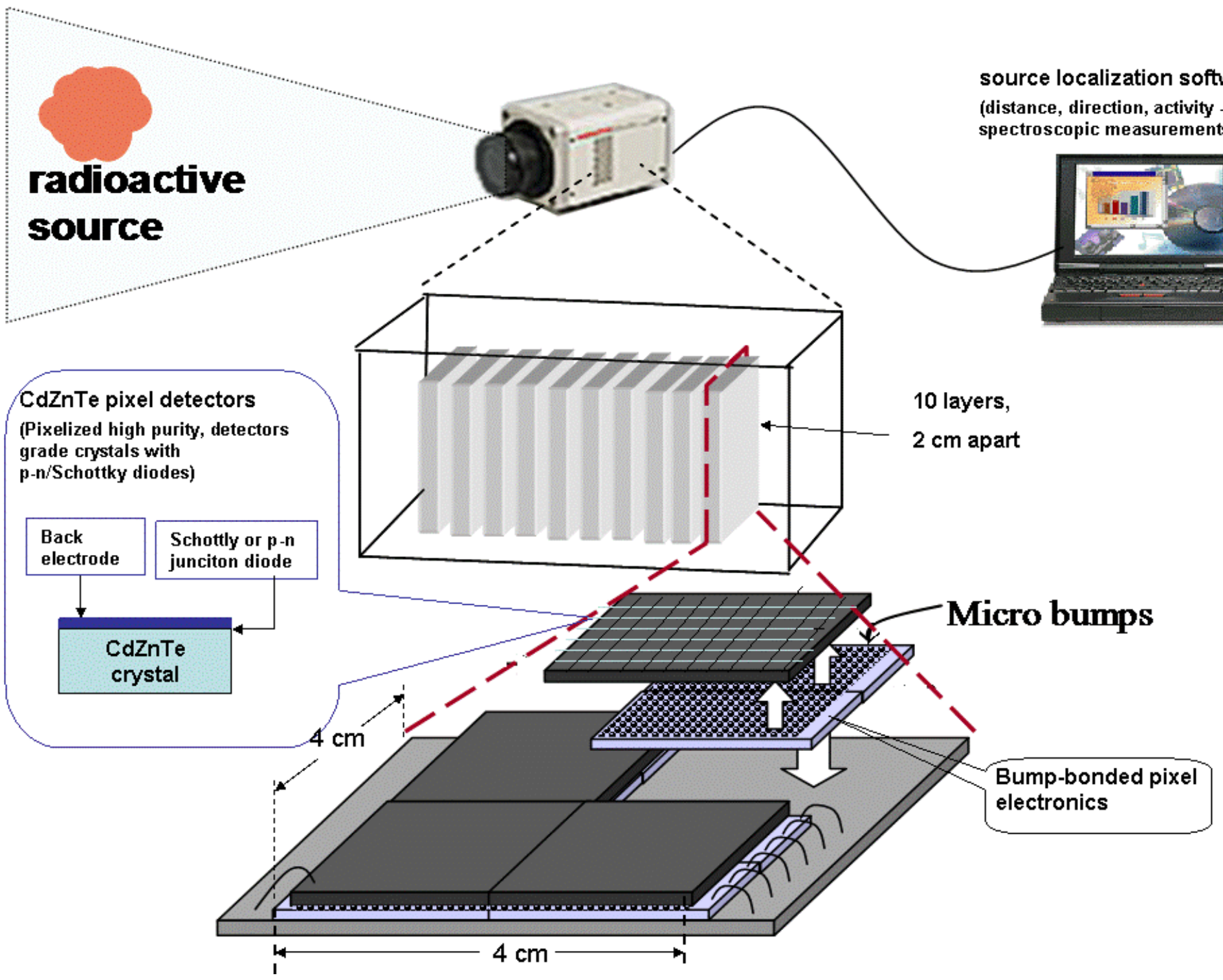

**FIGURE 1.** Schematic illustration of the COCAE instrument configuration

The proposed instrument consists of ten parallel planar layers, placed 2cm apart, and made of pixelated 2mm thick Cadmium Telluride (CdTe) crystals occupying an area of 4cmx4cm.

Each detector's layer consists of a two dimensional array of pixels (100x100) of 400μm pitch, bump-bonded on a two dimensional array of silicon readout CMOS circuits of 300μm thickness. Both pixels and readout arrays are on top of an $Al_2O_3$ supporting printed circuit board layer.

The most crucial parameters of the detector for the source localization are its detection efficiency and energy resolution, which influence the event statistics and the uncertainty of the Compton scattering angle determination

As to the detection efficiency, CdTe semiconductor crystals compared to Germanium (Ge) and Sodium Iodide (NaI) detectors have in principle higher efficiency due to the higher atomic number. In order to achieve even better efficiency, a thick CdTe detector of several mm would be needed, but an increase of the crystal thickness would deteriorate the detector energy resolution due to the effect of incomplete charge collection of CdTe semiconductors. To bypass this restriction COCAE instrument is designed as system of ten 2mm stacked detectors, instead of one thick mono-crystal.

As to the energy resolution, even not very high quality CdTe crystals exhibit better energy resolution than NaI detectors. On the other hand the Doppler broadening effect (see section C) is higher for CdTe detectors compared to NaI and Ge detectors. The challenge for the COCAE instrument is to achieve an energy resolution comparable to the one of high purity Germanium detectors, without the need of cryogenics (CdTe semiconductors can be operated at room temperature due to their high energy bandwidth).

## B. PRINCIPLE OF COMPTON RECONSTRUCTION

Instruments like COCAE that exploit the Compton imaging technique deduce the energy of the incident gamma ray photons as well as their origin within a cone, by measuring the energy depositions and the positions of the Compton scattering interactions recorded in the detector. A typical design of such an instrument consists of two types of detectors, the scatter detector with relatively low atomic number, where the Compton scattering occurs, and the absorber with relatively high atomic number, in which the scattered photon is ideally totally absorbed. The COCAE instrument's CdTe detectors work both as a scatter, thanks to their arrangement into thin layers, and as an absorber, due to the large atomic numbers of Cd (Z=48) and Te (Z=52), resulting into a high photo-absorption efficiency.

Figure 2 illustrates the schematic of the Compton reconstruction principle [1]. Considering a photon with energy $E_0$ incident on the detector's sensitive area that undergoes a Compton scattering, it will create a recoil electron of energy $E_e$, quickly absorbed and measured by the detector, and a scattered photon of energy $E_g$. The scattered photon ideally deposits its energy in the detector in a series of one or more interactions and is finally absorbed via a photoelectric interaction. The photon scatter angle φ is related to the measured energy depositions according to the Compton formula:

$$\cos\phi = 1 - m_0c^2\left(\frac{1}{E_g} - \frac{1}{E_g + E_e}\right) \quad (1)$$

where $m_0c^2$ is the rest energy of the electron.

In addition the initial gamma ray direction $\hat{r}_0$ (unit vector) is geometrically related to the photon scatter angle $\phi$:

$$\cos\phi = \hat{r}_0 \cdot \hat{u} \quad (2)$$

where

$$\hat{u} = \frac{\vec{r}_2 - \vec{r}_1}{\left|\vec{r}_2 - \vec{r}_1\right|} \quad (3)$$

and $\vec{r}_1$, $\vec{r}_2$ are the positions of the interactions.

By measuring the energy depositions ($E_e, E_g$) as well as the positions of the interactions ($\vec{r}_1$, $\vec{r}_2$), the incident gamma ray photon direction is restricted to a cone, the opening angle of which is the Compton scatter angle $\phi$.

Successive interactions of the emitted gamma rays create overlapping cones according to equation (2) and the source location is the intersection of all measured cones (Figure 2b). In principle, three cones are sufficient to reconstruct the image of a point source. In practice, due to measurement errors and incomplete photon absorption, a large number of reconstructed cones are needed to derive the source location accurately.

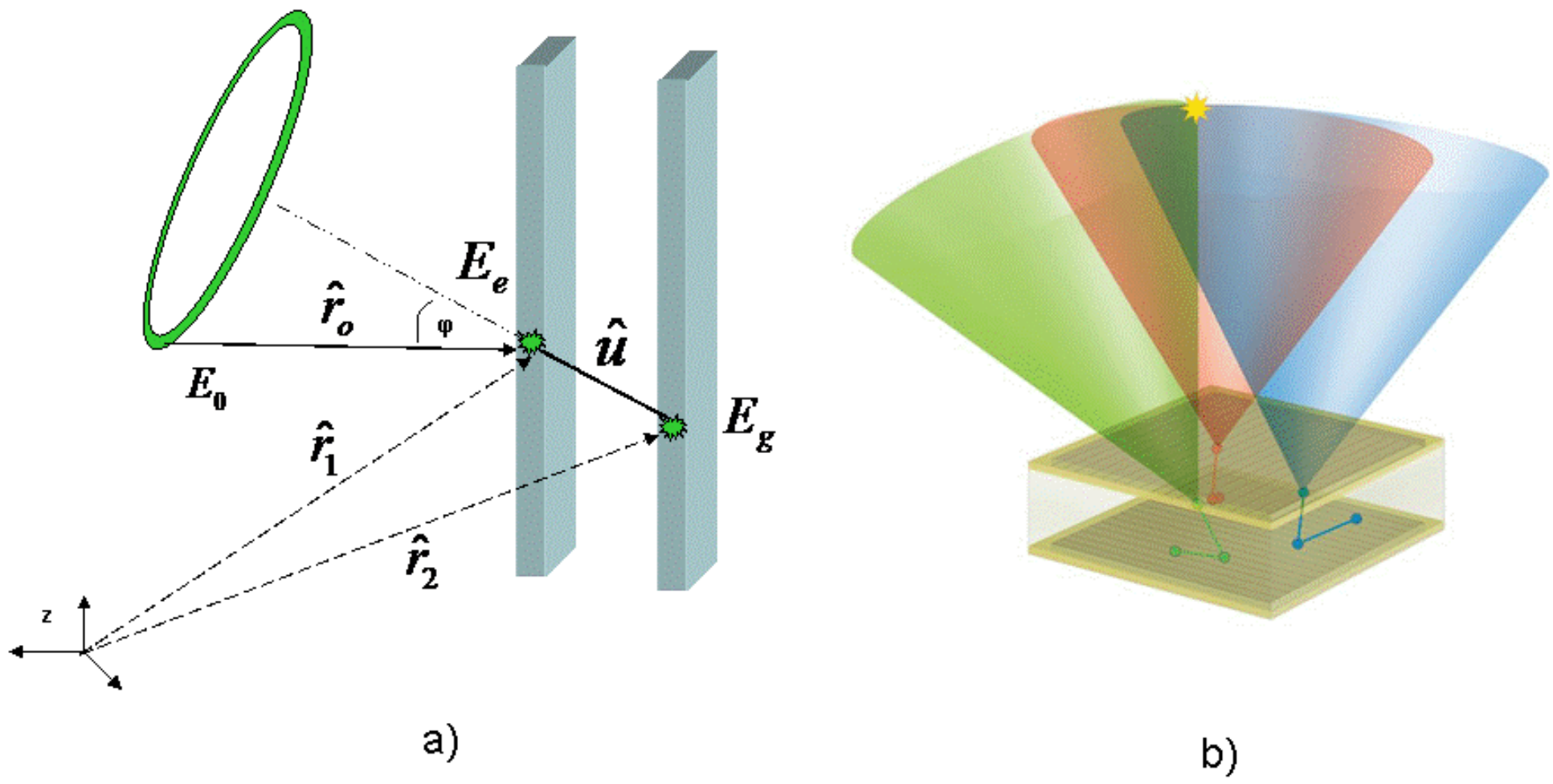

**FIGURE 2.** The principle of Compton reconstruction a) kinematics of Compton scattering b) Illustration of Compton imaging

The instrument's ability to localize radioactive sources accurately depends on the efficient reconstruction of the incident direction of the photons emitted by the source that interact through Compton scattering with the detector. Thus, the determination of the correct sequence of photon interactions in the detector is crucial. It has to be noted that in a real detector, an energy deposition is recorded only if its value is above a minimum threshold, therefore it is called "a hit". The importance of the determination of the correct sequence can be easily visualized in the case of a fully absorbed photon interacting in the active parts of the detector via a Compton scattering (dual hit event), as is illustrated in Figure 3.

Since the distance between the layers is very small in a portable instrument such as COCAE, it is impossible to have a timing tag for each hit, thus for a dual hit event, there are two possible hit orderings, each one resulting in a different direction for the initial photon.

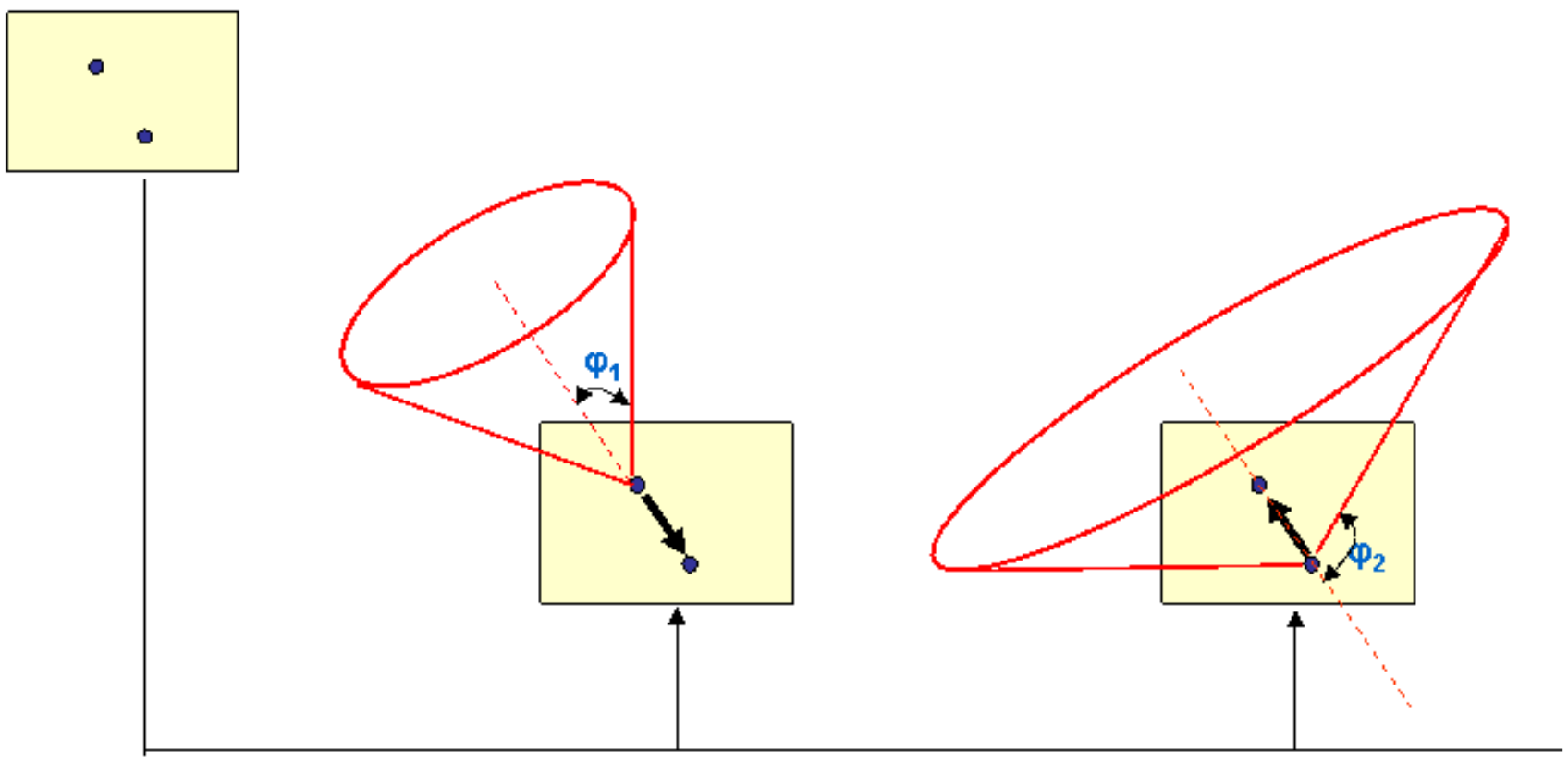

**FIGURE 3.** Compton sequence reconstruction of an event having two hits in the active detector's volume. An incorrect determination of the hit sequence results in an estimation of the Compton angle that could largely deviate from the correct value.

The algorithms for the Compton scattering sequence reconstruction aim at identifying the hit order in absence of timing information by exploiting the kinematical and geometrical information of the event as well as statistical considerations.

The main task of the present work is to compare the performance of different Compton sequence reconstruction algorithms, tested on a large number of Monte Carlo simulated events interacting with the model of the COCAE detector, in order to select the best performing one for the reconstruction procedure of a Compton event.

# C. MONTE CARLO SIMULATION

The COCAE instrument is modelled by an open-source object-oriented software library (MEGAlib [2]) which comprises detector geometry construction tools, reconstruction methods and high level data analysis tools. For the Monte Carlo studies, MEGAlib provides interface to the Geant4 [3] simulation toolkit. In Geant4, the Low Energy Compton Scattering (G4LECS [4]) package is used to accurately model the Compton scattering by including Compton cross sections modified for bound electron momentum, as well as corrections on the resulting changes in the scattered particle energies (known as Doppler broadening)[2].

The Monte Carlo simulation study encompasses the following steps:

A) At the first step, the exact COCAE detector geometry is modelled. Since Compton scatterings can occur in both sensitive and non-sensitive materials of the detector, so resulting into incomplete energy measurements, special care has been taken to incorporate an accurate geometric and physical description of the detector's passive materials. This is a precondition along with the implementation of the correct isotopic composition of all detectors' materials as well as of all the corresponding particle cross sections in order to ensure an accurate simulation of the real detectors' performance.

B) At the second step, a point gamma ray source with isotropic emission is modelled, emitting photons in an energy range from 60keV up to 2000keV. The gamma ray source is placed in front of the detector, on its axis of symmetry and at a distance of 40cm from the first detecting layer. A very large number of photons ($\sim 2 \times 10^9$) are generated and are allowed to interact with the COCAE model. For accurate simulation of the detector response the simulation is ensured to take into account all relevant physical processes (low energy electromagnetic processes including Doppler broadening).

The output of this simulation step is a collection of hits consisting of energies highly idealized and positions limited by the pixel's dimensions. In order to adapt the ideal simulated data to realistic energy measurements, the simulated photon energies are blurred according to Gaussian distributions with a FWHM that varies from 3.5% at energies around 100keV down to 1% at energies above 662keV, assumed to be in accordance with realistic energy measurements.

C) The third step of the simulation is the process of event reconstruction at which dedicated algorithms are used in order to group together the individual simulated hits into events and to identify their original interaction process. Thus, at the end of the event reconstruction process the data are represented by event types (e.g. Compton scattering or photo-effect event) and their associated information (e.g. energy and direction of the Compton scattered gamma ray and of the recoil electron).

The event reconstruction is split into two steps:

a) Clustering (blobbing adjacent hits into one larger hit, called from now on an Energy Cluster or simply a cluster), and

b) Compton sequence reconstruction (identifying the sequence of Compton interactions).

For the clustering, the energy of the eight closest neighbor pixels to a hit combine to form a cluster (Figure 4-a). The energy of the cluster is the total energy of its pixels and its position is calculated via an energy-based center of gravity [5]. Figure 4-b shows the average number of hits in a cluster, for fully absorbed events, as a function of the incident photon energy. As it is expected, the mean value of hits combined into a cluster increases with the energy, since at higher energies the scattered electrons travel longer distances in the detector's material, thus they activate more pixels.

---

[2] However, as for GEANT4.9.2 the G4LowEnergyCompton physics package includes a treatment of Doppler broadening very similar to that provide by G4LECS package.

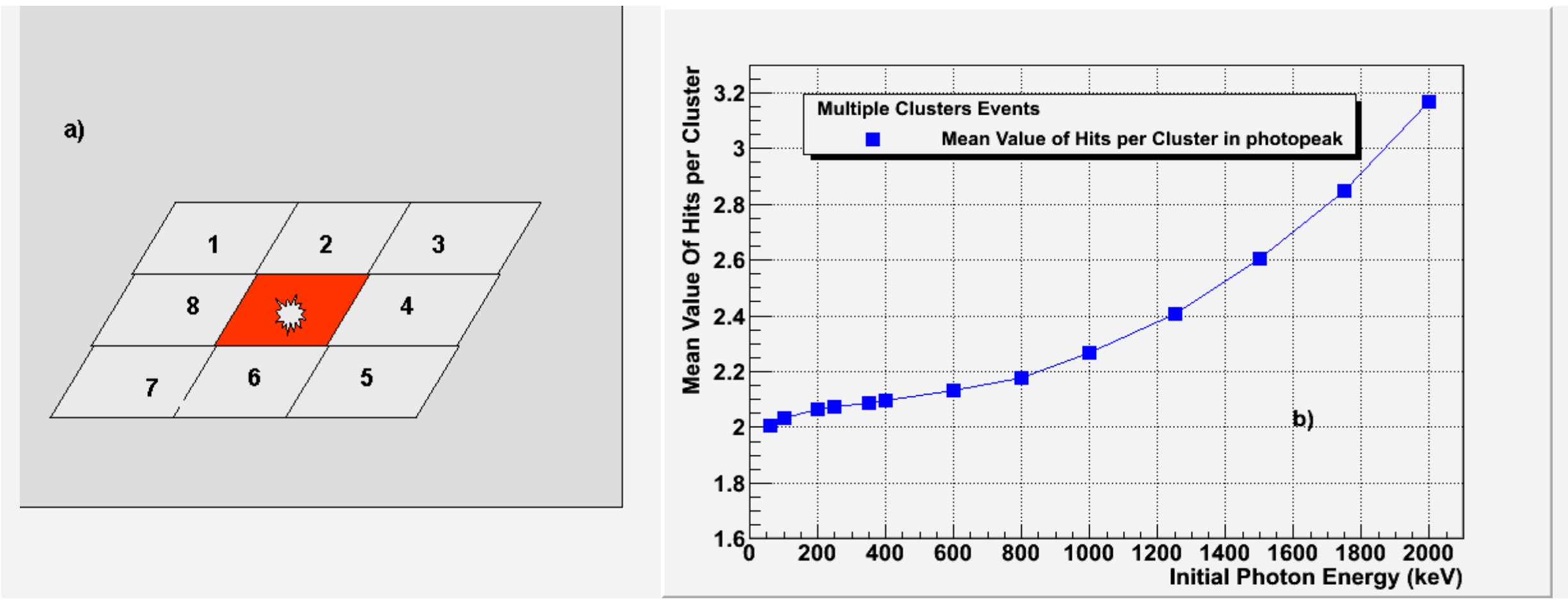

**FIGURE 4.** a) Clustering adjacent hits into a larger hit b) Average number of hits per cluster, as a function of the incident photon energy

For the sequence reconstruction, the clusters must be sorted in the order in which the interactions of the original particles occurred inside the detector. Since the distance between successive interactions is short, time of flight measurements are not possible. Thus, the interaction sequence cannot be determined directly. Consequently, if N clusters are recorded in the detector, there are N! possible sequences, of which only one is the correct one, whereas the rest form a combinatorial background. The objective of the sequence reconstruction algorithms is to select the most probable sequence among all possible ones and to eliminate the combinatorial.

In order to achieve efficient reconstruction of events a sufficient understanding of the gamma ray interactions with the detector's materials is necessary. Thus, simulation studies have been performed to understand the detector response under different radiation fields.

In the sub-MeV to MeV energy range, the four dominating interactions of photons with matter are the photoelectric absorption, the Compton scattering, the pair production and the Rayleigh scattering mechanisms. The shape of the spectrum of the energy deposited on the detector depends largely on the mechanism via which the incident photon primarily interacts with the detector. Figure 5 shows the simulated relative contribution of the photon's first interaction in the detector's sensitive areas as a function of the incident photon energy. It can be noticed that the probability that the first interaction of the incident photon is a photoelectric effect dominates at low energies whereas the probability that the first interaction is a Compton scattering dominates for photons more energetic than 200keV (~30%), reaching ~90% at energies above 800keV.

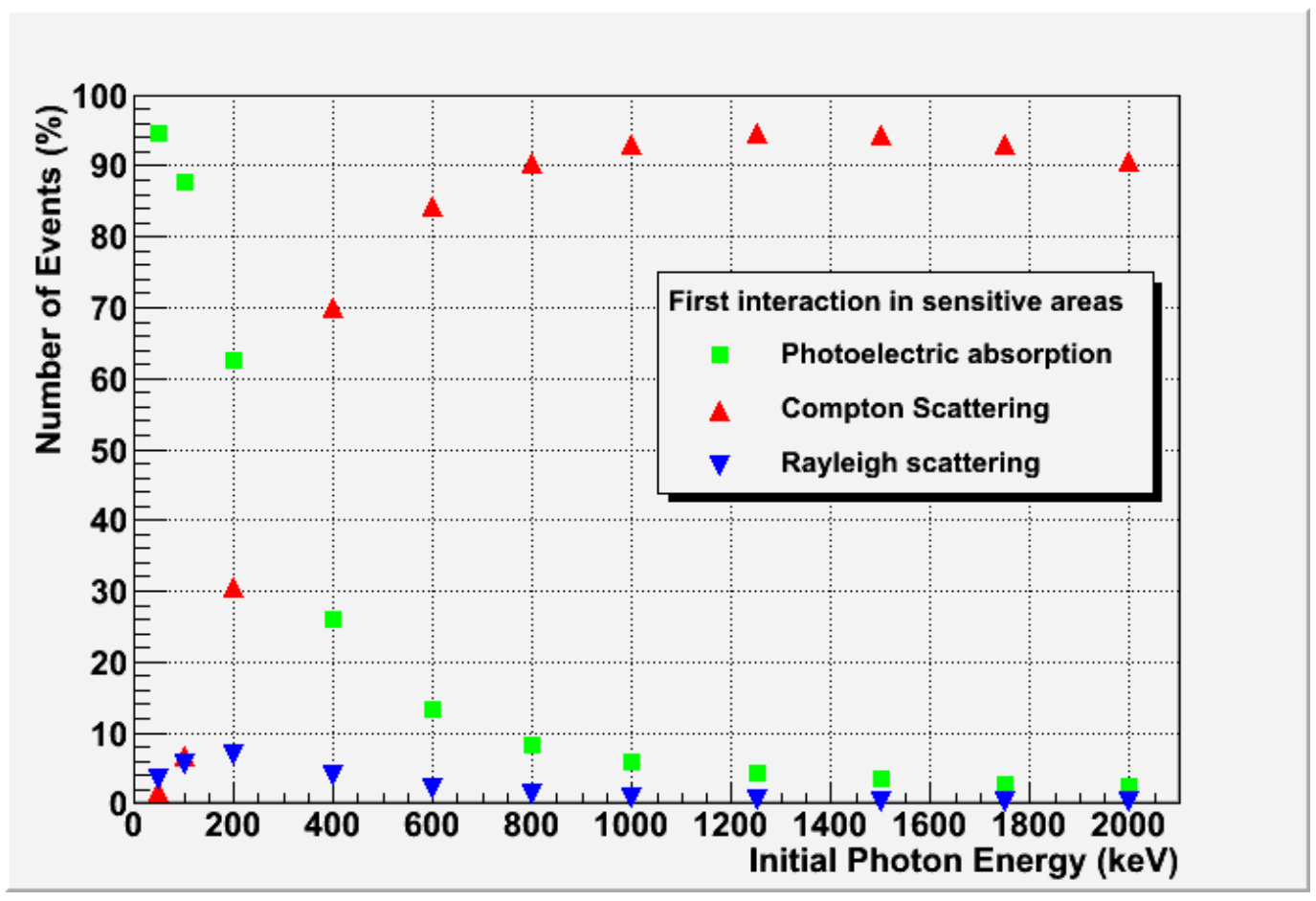

**FIGURE 5.** Relative contribution of the photon's first interaction in the detector's sensitive areas as a function of the incident photon energy.

Shown in Figure 6 is the spectrum of the simulated energy deposition on the detector's active volume, of different mono-energetic incident gamma ray energies varying from 200keV up to 1500keV.

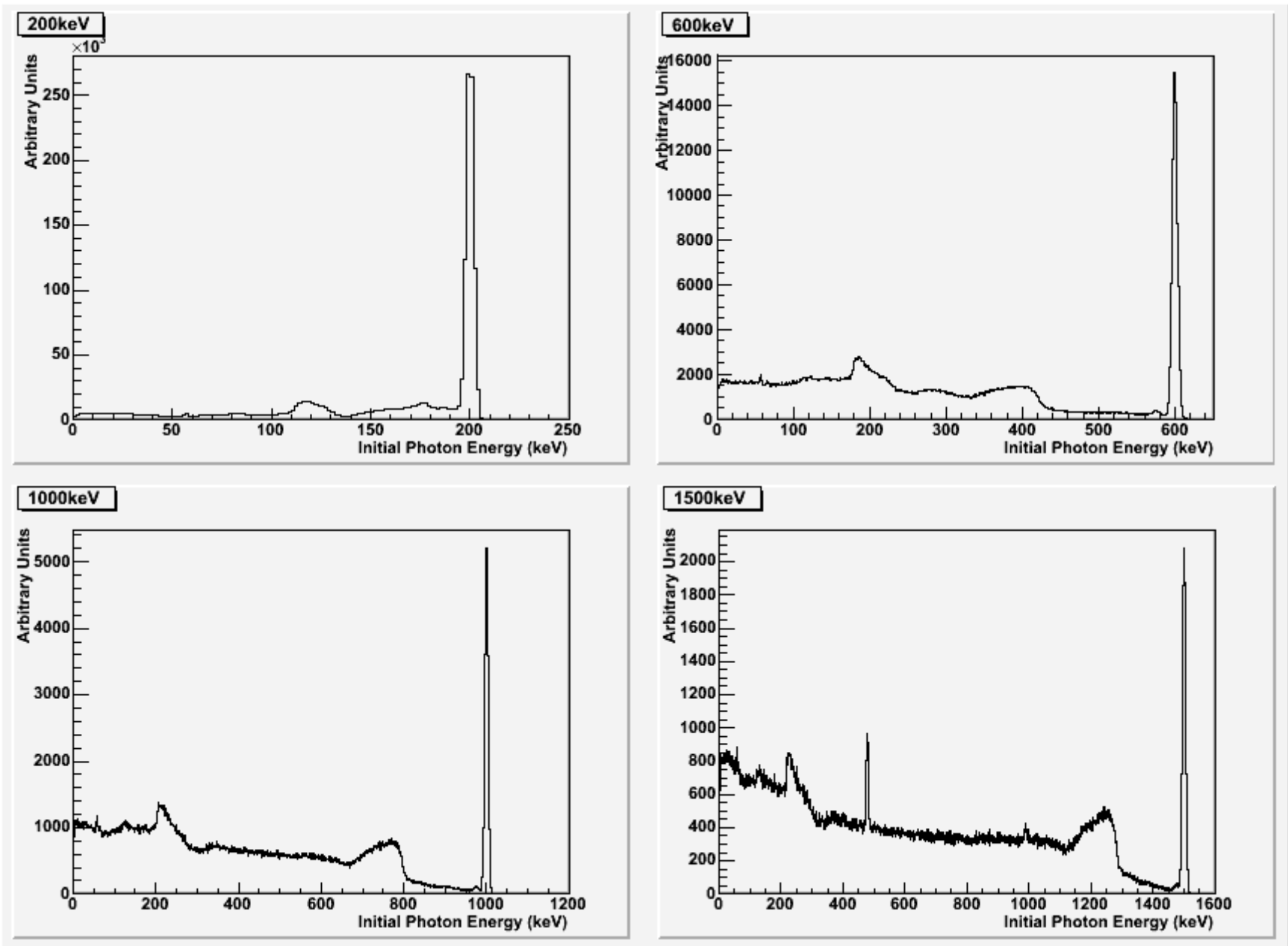

**FIGURE 6.** Typical energy deposition spectra for various incident gamma rays energies illustrating the photo-peak and the Compton plateau.

Interactions via a photoelectric effect result in full energy absorption and contribute to the full energy peak (photo-peak). In the Compton scattering interactions, the scattered electron carries a fraction of the initial gamma ray photon, whereas the scattered photon is either absorbed or escapes the detector**.** Figure 7 shows the ratio of events that are completely absorbed in the detector over the total number of events that interact with the detector's active volume and give a measured energy deposition.

Notice that in the following studies the events are considered to be fully absorbed when the photon's energy is within $\pm 3\sigma$ from the incident photon's energy.

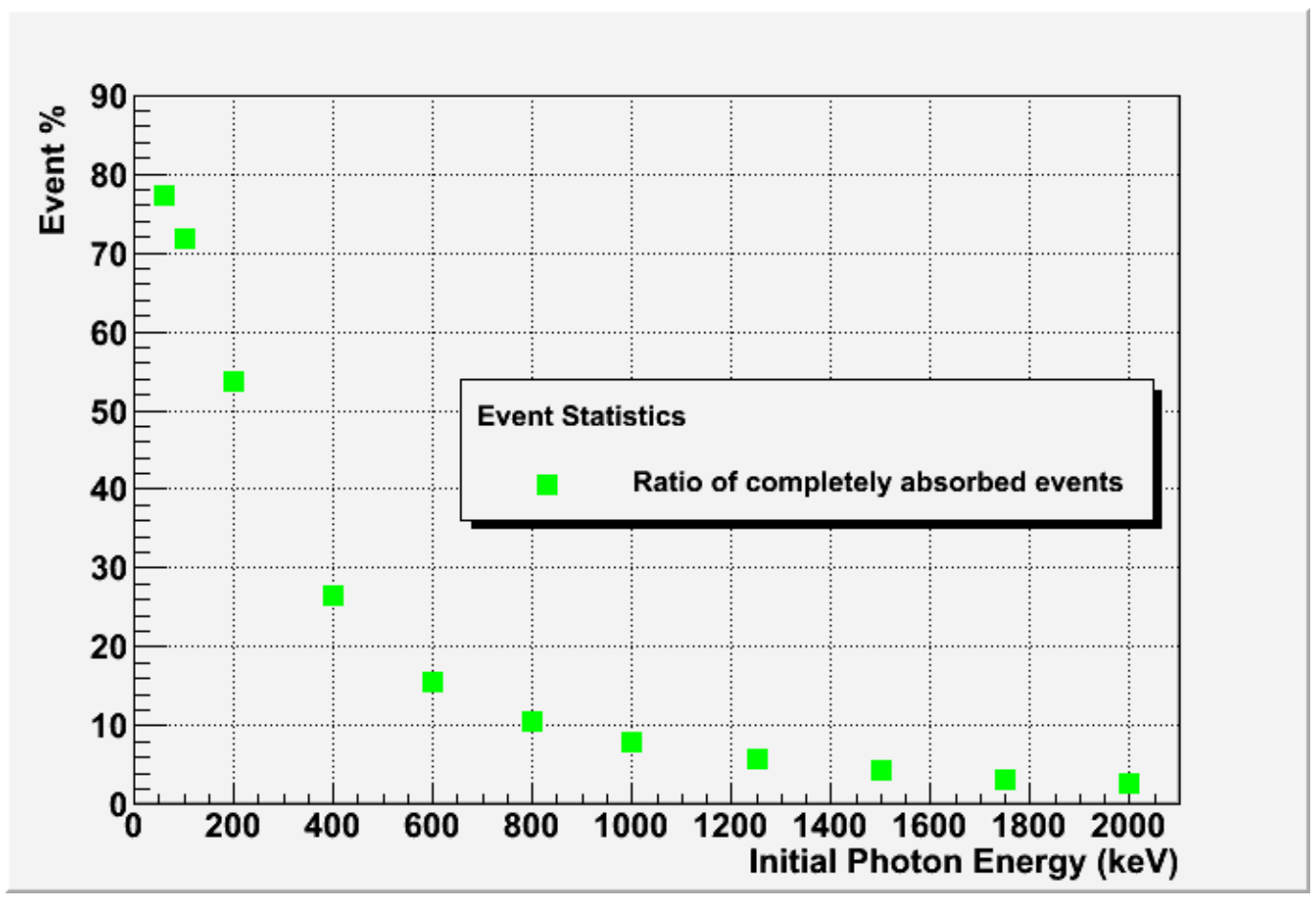

**FIGURE 7.** Percentage of fully absorbed events over the total number of events having energy depositions in the detector's active volume, as a function of the initial photon's energy.

In the energy spectrum (Figure 6), the full energy peak originates from two different types of events: a) events having one cluster in the detector; these are events for which the total energy of the photon is transferred to the detector's electrons at once due to photoelectric effect, and b) multiple cluster events for which the total energy of the photon is transferred to the detector's electrons after a sequence of Compton scatterings followed by a photoelectric interaction. It is evident that only Compton scattering events that are fully absorbed are useful for the Compton scattering reconstruction. Figure 8 summarizes the simulation results for the contribution to the photo-peak part of the energy spectrum of events that have multiple clusters (mainly events that undergo multiple Compton scattering, which in turn gets photo-absorbed).

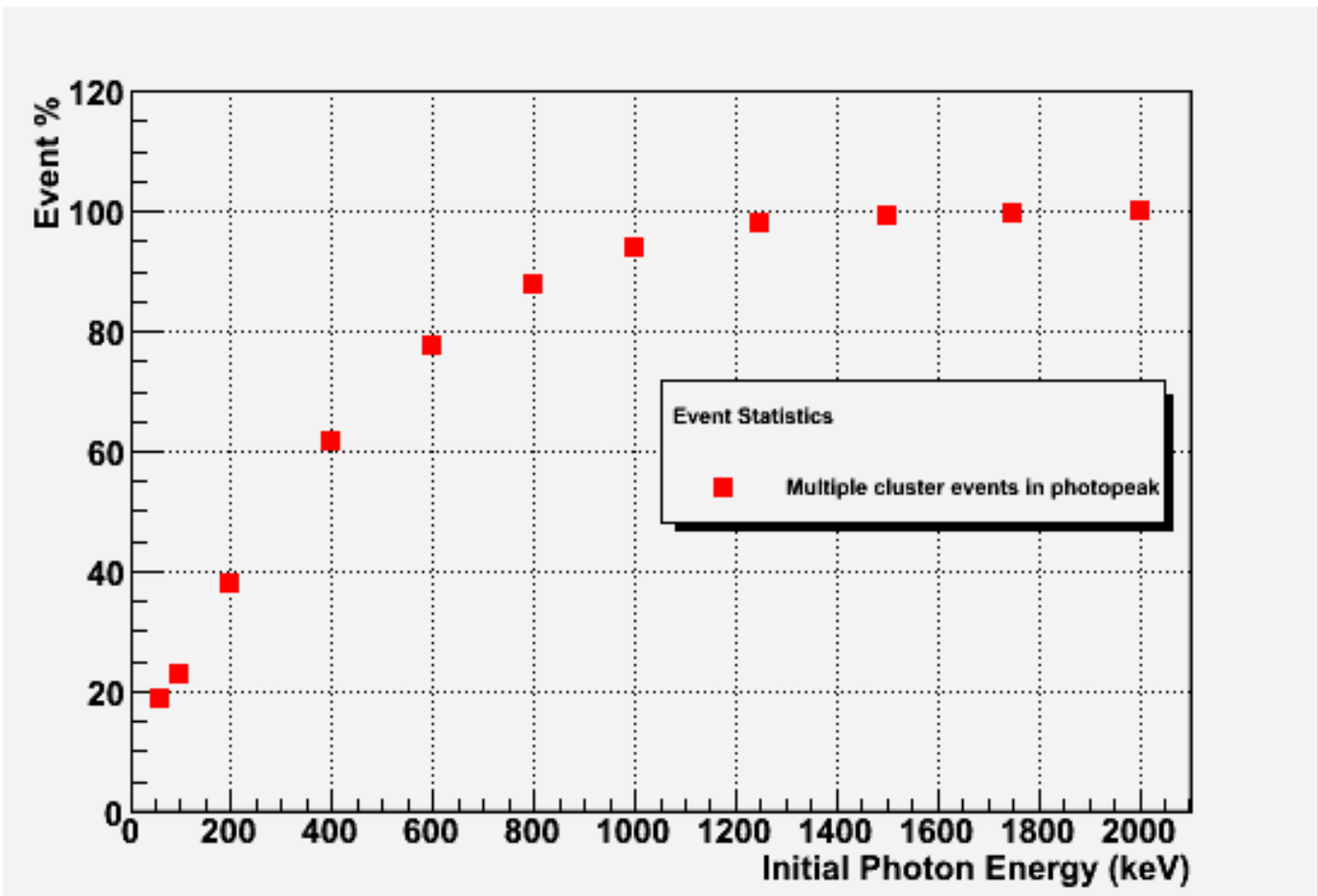

**FIGURE 8.** Percentage of fully absorbed multiple cluster events as a function of the incident gamma-ray energy.

Furthermore, it has been verified that the average number of Compton interactions (events having more than one cluster) contributing to the photo-peak increases as the incident photon energy increases (Figure 9). The available information from multiple cluster events is crucial for the reconstruction of the initial photon direction, thus the number of events suitable for reconstruction becomes higher as the photon's initial energy increases.

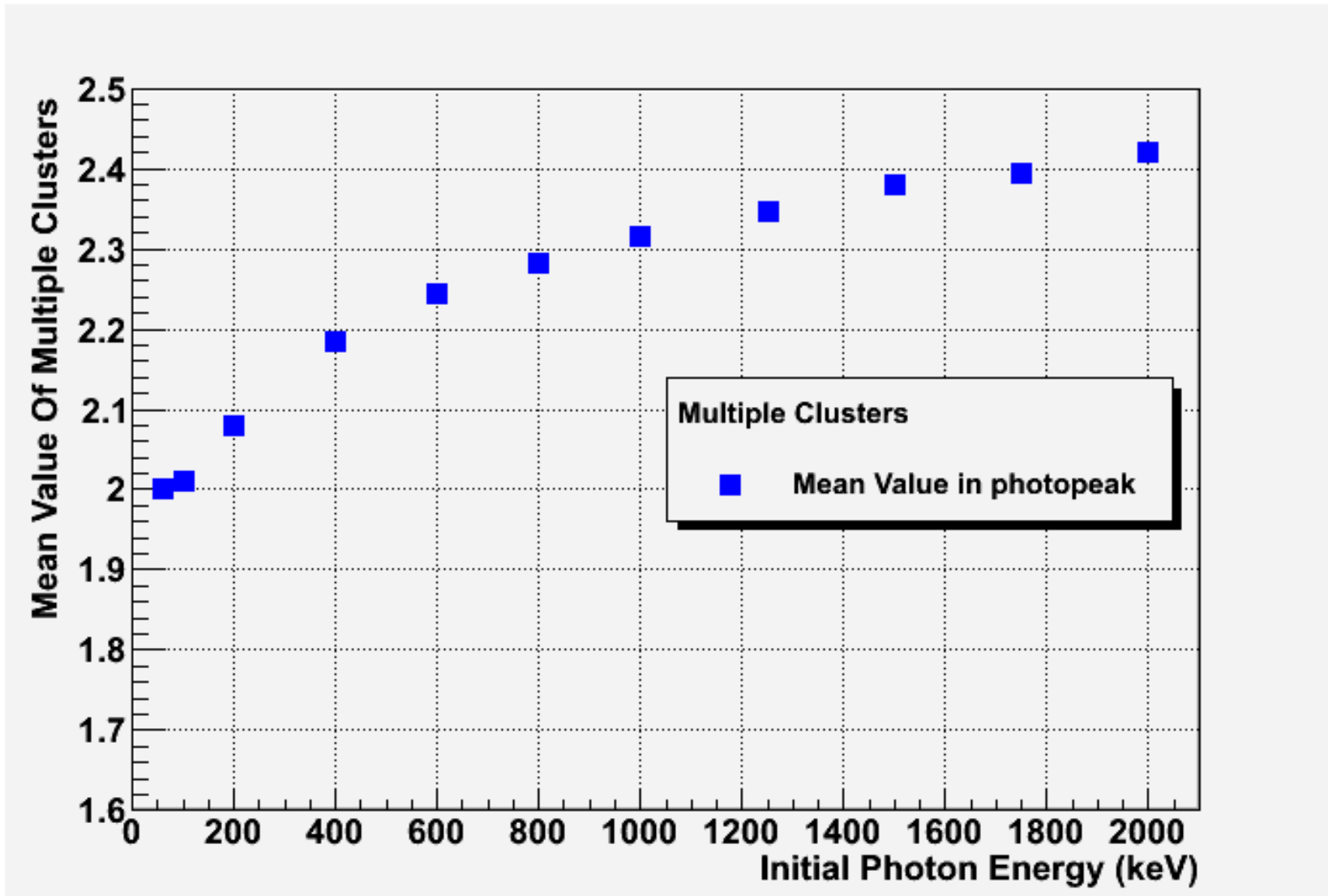

**FIGURE 9.** The mean value of clusters of multiple-cluster events totally absorbed, as a function of the incident photon energy.

However, the complexity of the sequence reconstruction algorithms goes with the factorial of the number of interactions (see section D2), thus reconstructing events with more than seven clusters is getting difficult. Therefore, such events are rejected from our study. Fortunately, the number of the rejected events is extremely limited for energies up to 2000keV, as indicated in Figure 9, where the highest value of the average of multiple clusters in the totally absorbed events is just over 2.4.

## D. SEQUENCE COMPTON RECONSTRUCTION

As it has been mentioned in the introduction, in order to reconstruct the Compton scattering events accurately, the order of the photon interactions in the COCAE instrument must be correctly determined.

Two different techniques for the sequence reconstruction have been exploited: the Dual Cluster Sequence reconstruction (DCS) applied on events with only two interactions in the detector's active volume and the Multiple Cluster Sequence reconstruction (MCS) applied on events with more than two interactions.

Figure 10 shows the percentage of fully absorbed events having two or more clusters in the detector, as a function of the initial photon energy. It is evident that the case of two-cluster events is predominant at all energies. The case of events having more than two clusters becomes more frequent in higher energies, reaching about ~40%, at photon energies of 2MeV.

Notice that in the following plots all errors are of the order of 1%, thus they are not visible.

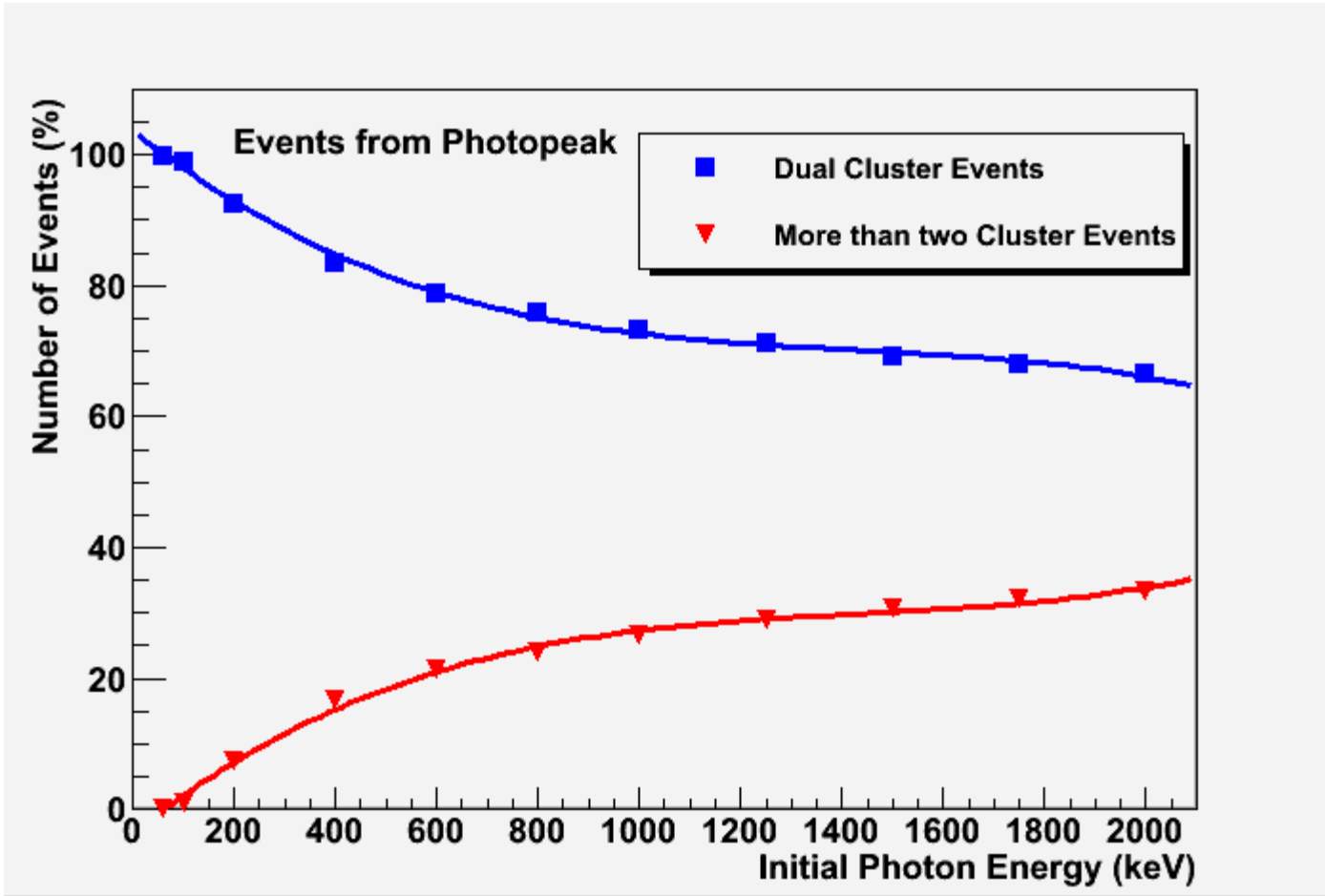

**FIGURE 10.** Photo-peak events: Percentage of two-cluster and more than two cluster events as a function of the incident gamma-rays energy.

Presented in the following sections are the different algorithms for the sequence reconstruction, extensively studied using a very large number (~2x$10^9$) of simulated events interacting with the model of the COCAE detector. The flow-cart of the reconstruction algorithms used is depicted in Figure 11.

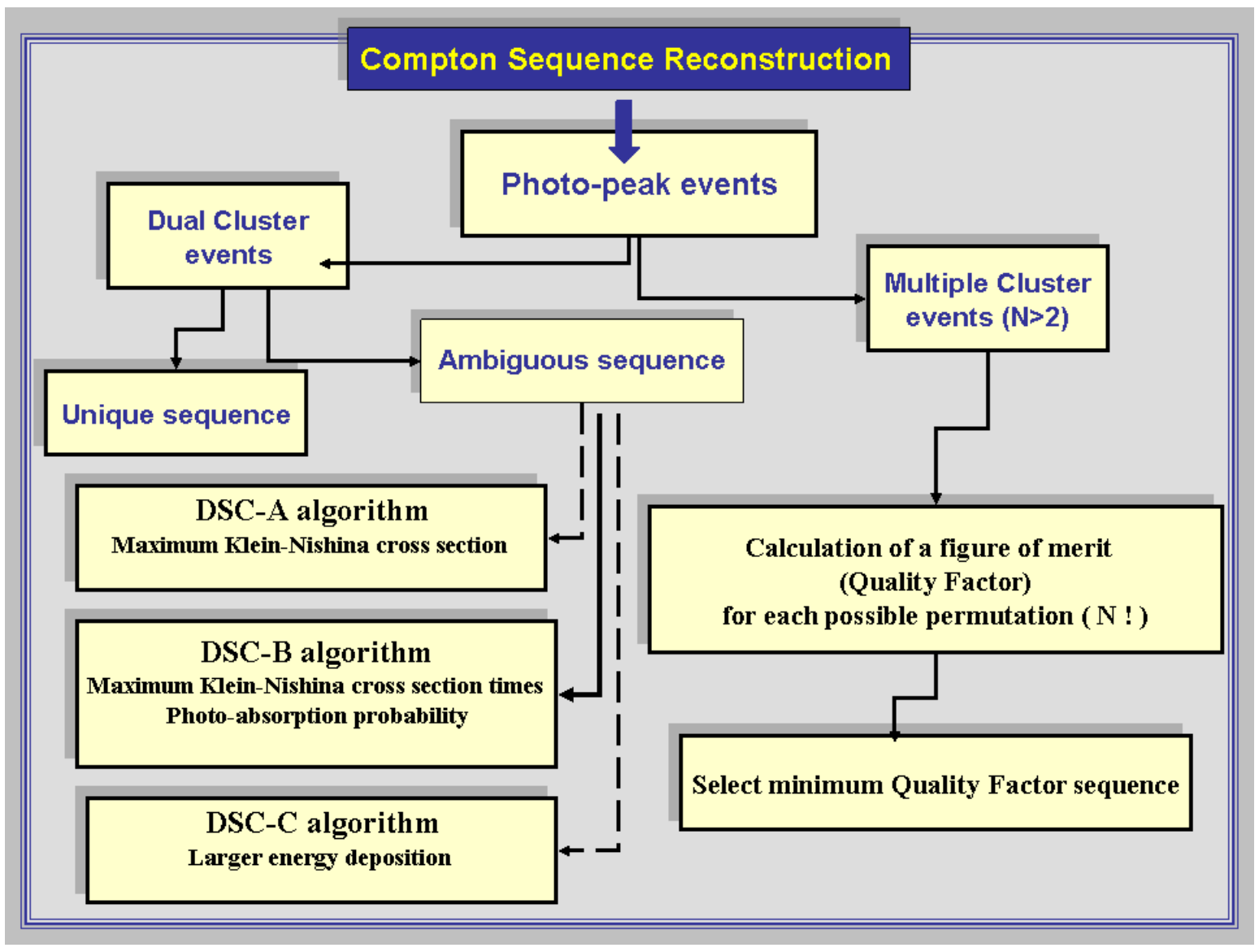

**FIGURE11.** Compton sequence reconstruction algorithms layout.

## D1. Dual Cluster Sequence reconstruction (DCS)

The first step of the sequence reconstruction of dual cluster Compton events is to apply the Compton kinematics in order to reject the non-physical sequences, i.e. to test whether the possible cluster sequences are compatible with the requirement that $\cos\phi \leq 1$ (equation (1)), namely, whether the electron ( $E_e$ ) and photon ( $E_g$ ) energies satisfy the following constrains, respectively:

$$\frac{m_o c^2 E_0}{2m_o c^2 + E_o} < E_g < E_0$$
$$0 < E_e < \frac{m_o c^2 E_0^2}{2m_o c^2 + E_o} \quad (4)$$

where $m_0 c^2$ is the rest energy of the electron and $E_0$ the energy of the initial gamma ray.

Figure 12 presents the fraction of fully absorbed two-cluster events which have a unique or ambiguous ordering, as a function of the incident photon energy, along with the fraction of the events for which both sequences are not allowed by Compton kinematics. It is evident that for incident photon energies below 200keV all of the dual cluster Compton events in the photo-peak have a unique ordering. It can also be noticed that for photon energies below 100keV both sequences are restricted. This can be attributed to the fact that at low energies, the two clusters originate from X-rays produced during the de-excitation of Cd and Te atoms internal excitation, emitted in any direction and interacting with the detector, thus they do not obey the Compton kinematics.

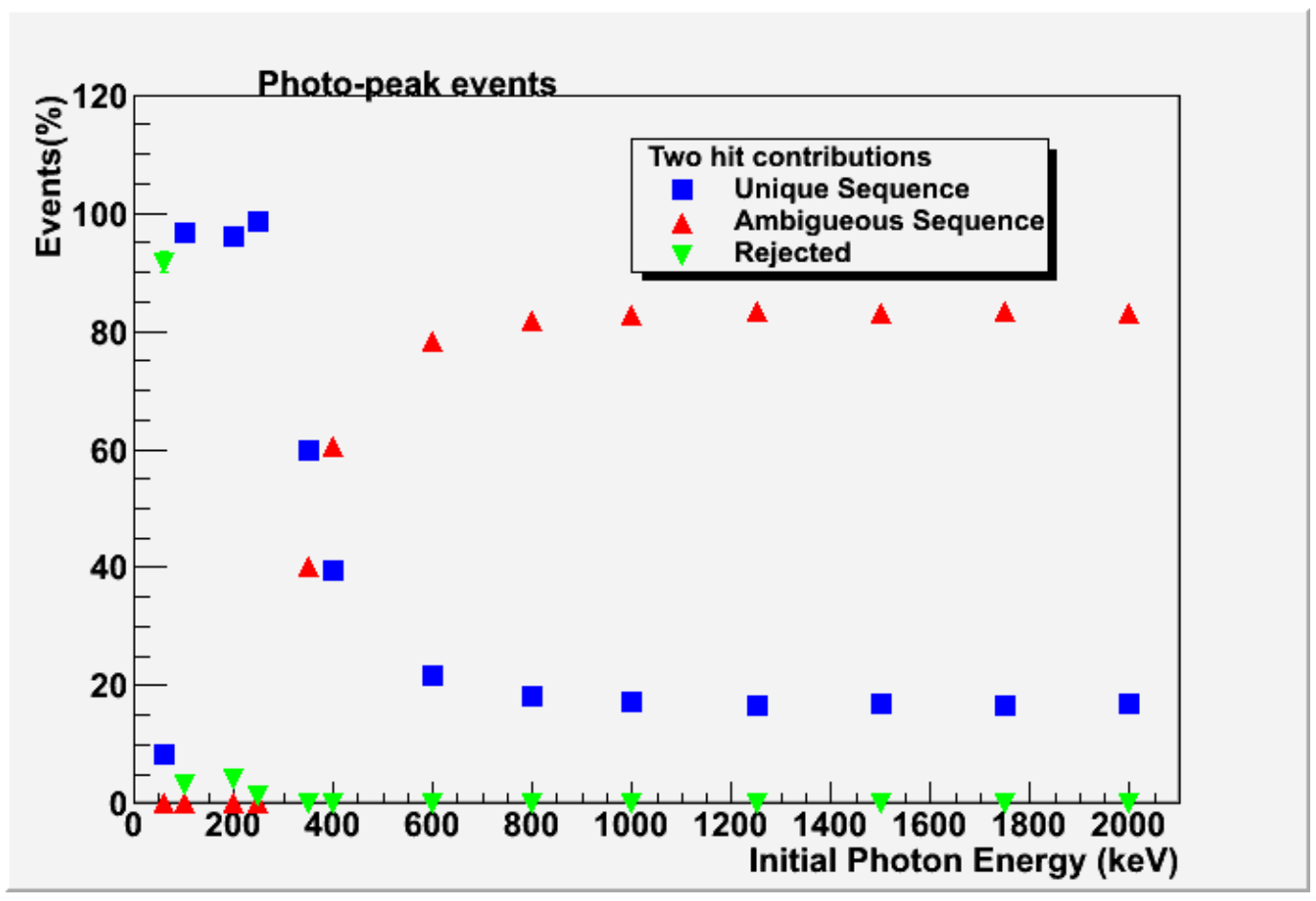

**FIGURE 12.** Dual cluster Compton events in the photo-peak as a function of the incident photon energy. The percentage of events with a unique possible ordering as well as those with ambiguous ordering is shown. Also shown is the fraction of events for which both orderings are restricted.

Furthermore, the photo-peak events having a unique sequence have been further studied in order to investigate whether their unique sequence is indeed the true sequence of the interactions (Figure 13). The true sequence has been derived directly from the Monte Carlo simulation where it is known a-priori.

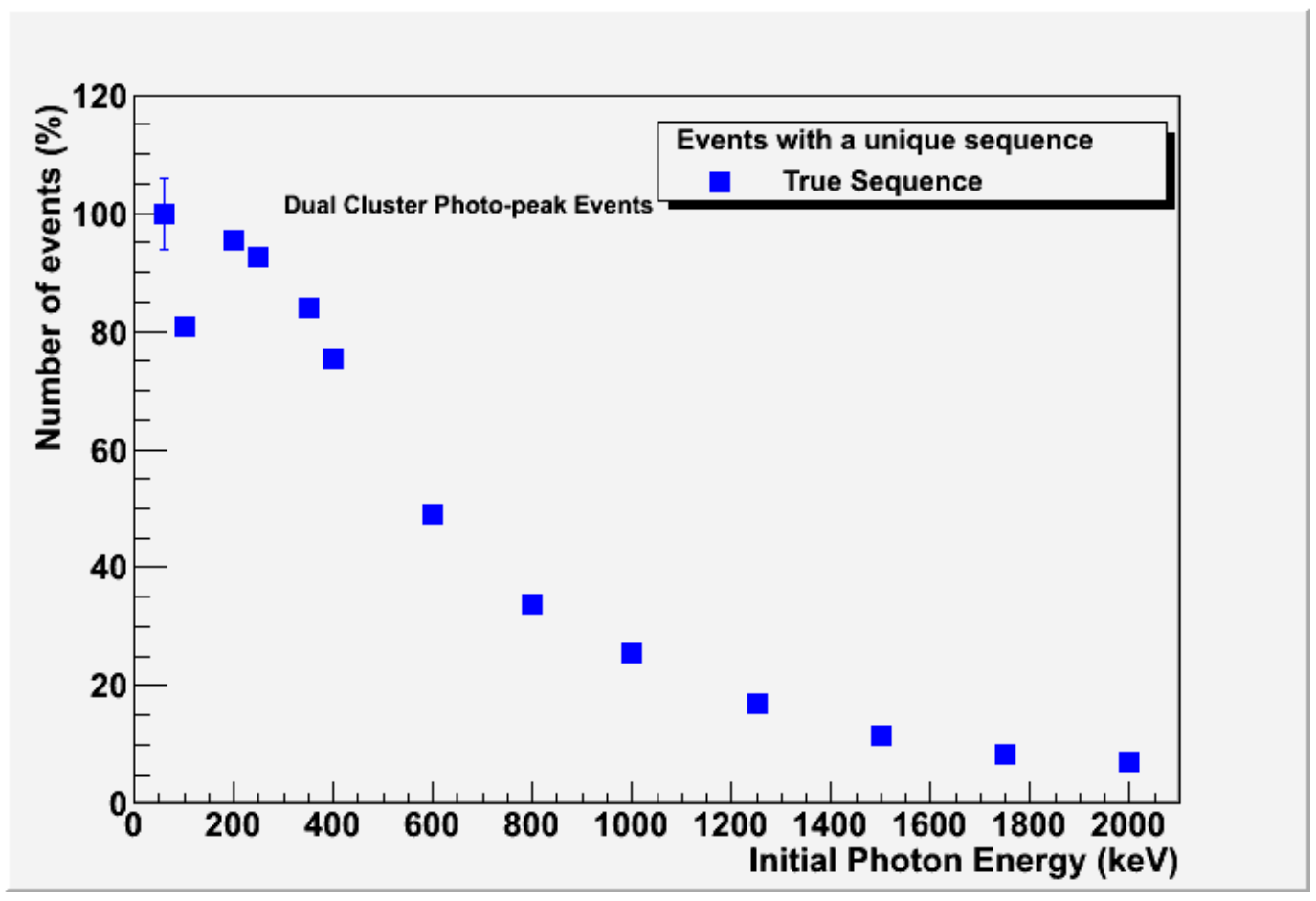

FIGURE 13. Two-cluster events for which their unique sequence compatible with Compton kinematics is indeed the true sequence of interactions.

For handling the ambiguous ordering events, three algorithms have been evaluated [5]:

A) According to the DSC-A algorithm, the Klein-Nishina differential Compton cross-section $\frac{d\sigma}{d\Omega}$) for un-polarized photons scattering off unbound electrons, is calculated for each of the two possible cluster sequences:

$$\frac{d\sigma}{d\Omega} = \frac{r_e^2}{2}\left(\frac{E_g}{E_0}\right)^2\left(\frac{E_g}{E_0} + \frac{E_0}{E_g} - \sin^2\phi\right) \tag{5}$$

where $r_e$ is the classical electron radius.

The sequence with the higher Klein-Nishina cross-section is assumed to be the correct one.

B) DCS-B algorithm calculates for the Klein-Nishina differential cross-section multiplied to the probability for absorption via a photoelectric effect and assumes that the sequence with the higher product probability is the correct one.

C) Finally according to DCS-C algorithm, the cluster that has the larger energy deposition is assumed to be the first Compton scattering.

Figure 14 shows the fraction of dual Compton events in the photo-peak that are correctly ordered, by applying the DCS-A, DCS-B and DCS-C algorithms, as a function of the incident photon energy.

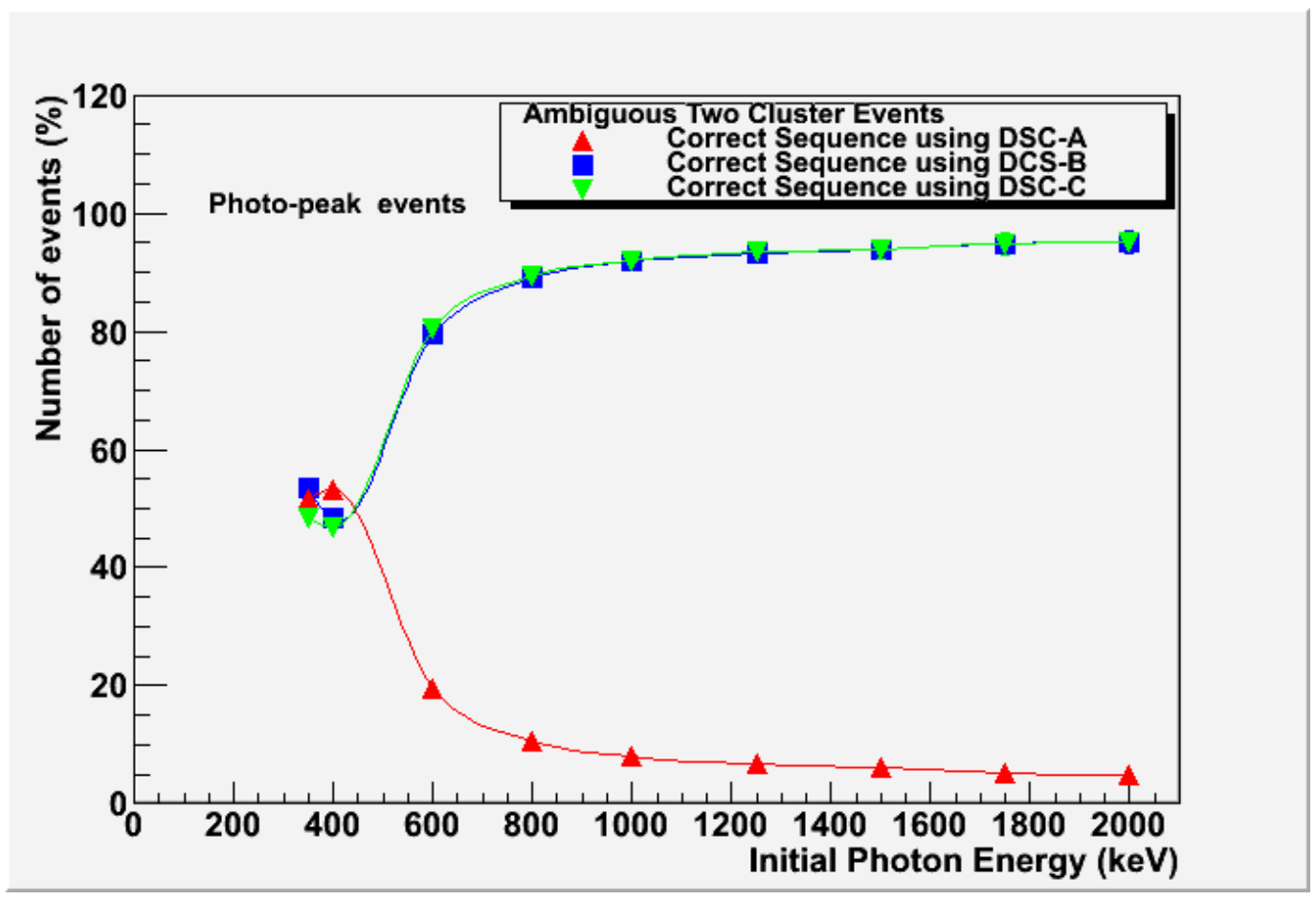

**FIGURE 14.** Compton sequence reconstruction efficiency for dual cluster Compton events in the photo-peak as a function of the incident photon for different reconstruction algorithms.

It can be noticed that the algorithms CDS-B and DCS-C similar performance, being able to identify the correct Compton sequence with an efficiency of about 95% for incident gamma energies above 800keV.

## D2. Multiple Cluster Sequence reconstruction (MCS)

For the sequence reconstruction of events with multiple clusters (at least two Compton interactions and one final complete absorption via a photoelectric effect), a technique has been proposed in [5], according to which a figure of merit (quality factor, QF) based on a generalized $\chi^2$ approach, is calculated for each one of the possible N! sequences, according to the following equation :

$$QF = \sum_{i=2}^{N-1} \frac{\left(\cos\phi_i^{kin} - \cos\phi_i^{geo}\right)^2}{\sigma^2_{\cos\phi_i^{kin}} + \sigma^2_{\cos\phi_i^{geo}}} \qquad (6)$$

where $\phi_i^{kin}$ is the Compton scatter angle at the $i^{th}$ scattering, calculated by the measured energy depositions according to a formula similar to equation (1), whereas $\phi_i^{geo}$ is the Compton scatter angle at the $i^{th}$ scattering calculated by the measured directions of the photons before and after the $i^{th}$ scattering, according to a formula similar to equation (2). The measurement errors $\sigma^2_{\cos\phi_i^{kin}}$ and $\sigma^2_{\cos\phi_i^{geo}}$ are due to the detector's energy and spatial resolution respectively.

Ideally the quality factor equals zero for the correct cluster sequence of Compton events, in the case where the photon is fully absorbed. Although measurement errors result in a quality factor greater than zero, the correct sequence is still most likely to correspond to its minimum value.

The distribution of the QF value is shown in Figure 15-a, for an incident photon with energy 600keV. Three cases are depicted in the same plot: the QF distribution for the false sequences, for the true sequences and for the sequences having the lowest QF value. It can be noticed that the efficiency of the MCS algorithm can be improved by applying upper thresholds on the value of QF (Figure 15-b), and by demanding that the distance between the first and the second interaction (known as lever arm) is more than 2cm (that is, the first two interactions are not in the same detecting layer).

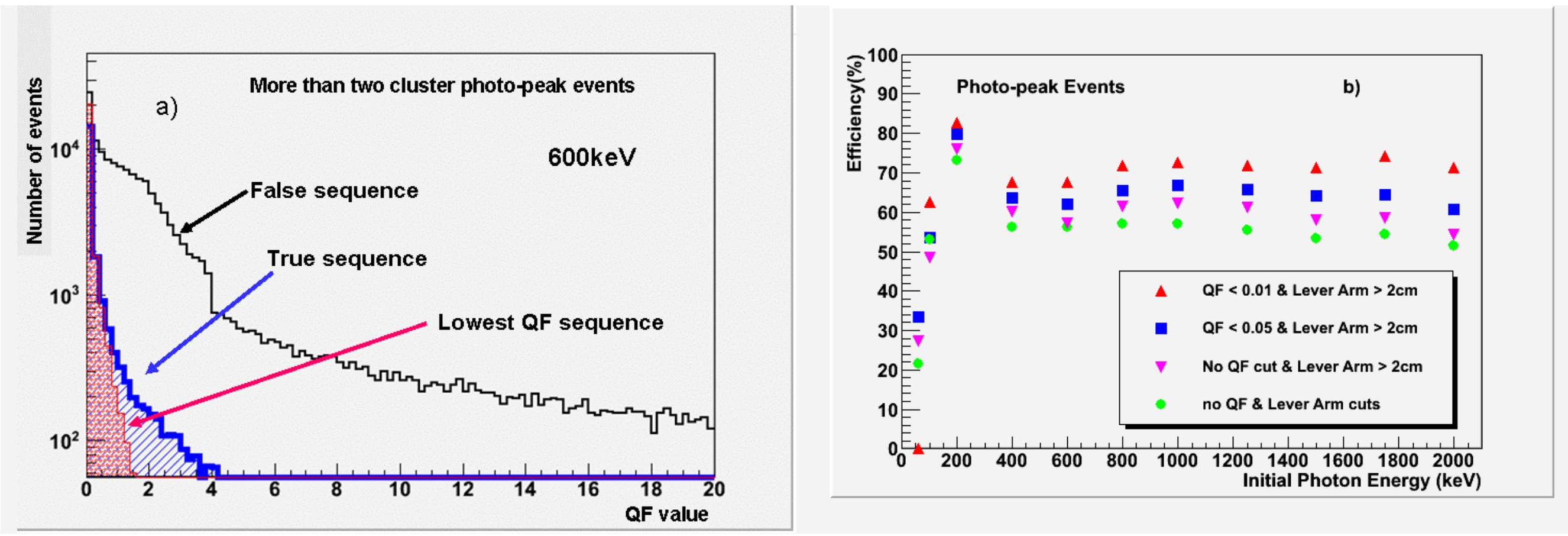

**FIGURE 15.** **a)** Multiple cluster photo-peak events at 600keV: QF distributions for the false and the true sequence of interactions, as well as for the interactions which have the lowest QF value. **b)** Efficiency as a function of initial photon energies, for different upper thresholds on the QF value and on the minimum distance between the two first interactions (lever arm).

It can be noticed that an upper threshold on the QF value improves the MCS algorithm's efficiency on the cost of a lower event statistics. Figure 16-a depicts both the efficiency and the event reduction as a function of the QF cut value, for multiple cluster photo-peak events with lever arm above 2 cm. For instance, only ~40% of the events have a QF value less than 0.01 (Figure 16-b)).

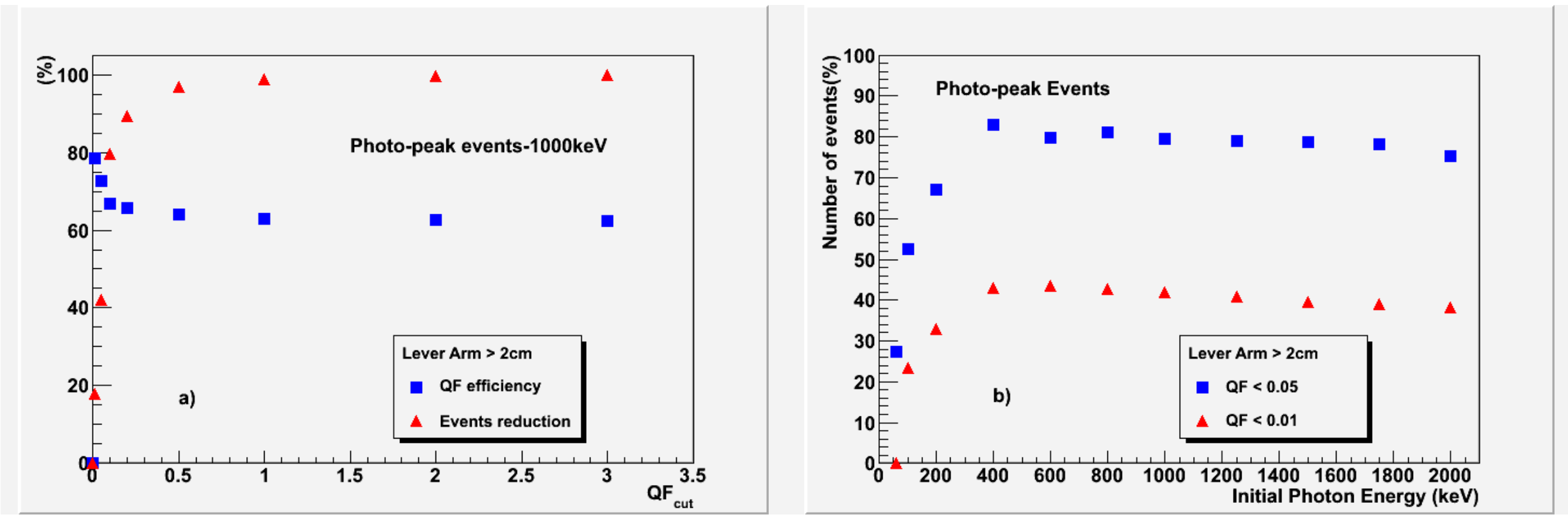

**FIGURE 16.** Photo-peak events having a lever-arm above 2cm: **a)** Multiple cluster photo-peak events at 1000keV: Efficiency and percentage of events reduction as a function of an upper threshold on the QF value. **b)** Percentage of events reduction as a function of initial photon's energy for different upper thresholds on the QF value.

## CONCLUSIONS

In the current study the performance of different algorithms for the Compton Sequence reconstruction has been assessed, using a large number of simulated events interacting with the model of the COCAE detector.

Specifically, two different such algorithms have been exploited: the Dual Cluster Sequence reconstruction (DCS) and the Multiple Cluster Sequence reconstruction (MCS). Among the DCS algorithms, the DCS-B and DCS-C have been proved to have similar performance, varying from ~50% up to ~90% (Figure 14). On the other hand the MCS algorithm correctly reconstructs the Compton sequences of more than two cluster events with an efficiency reaching ~55% at energies above 600keV (Figure 15-b). The MCS algorithm's performance can be improved up to ~70% by implying additional constraints on the minimum QF value and on the lever arm parameter, on the cost of lower event statistics.

The overall efficiency for the Compton sequence reconstruction including both the dual, as well as the multiple cluster fully absorbed events, using the DCS-B algorithm and the MCS without any constrain respectively, is summarized in Figure 17.

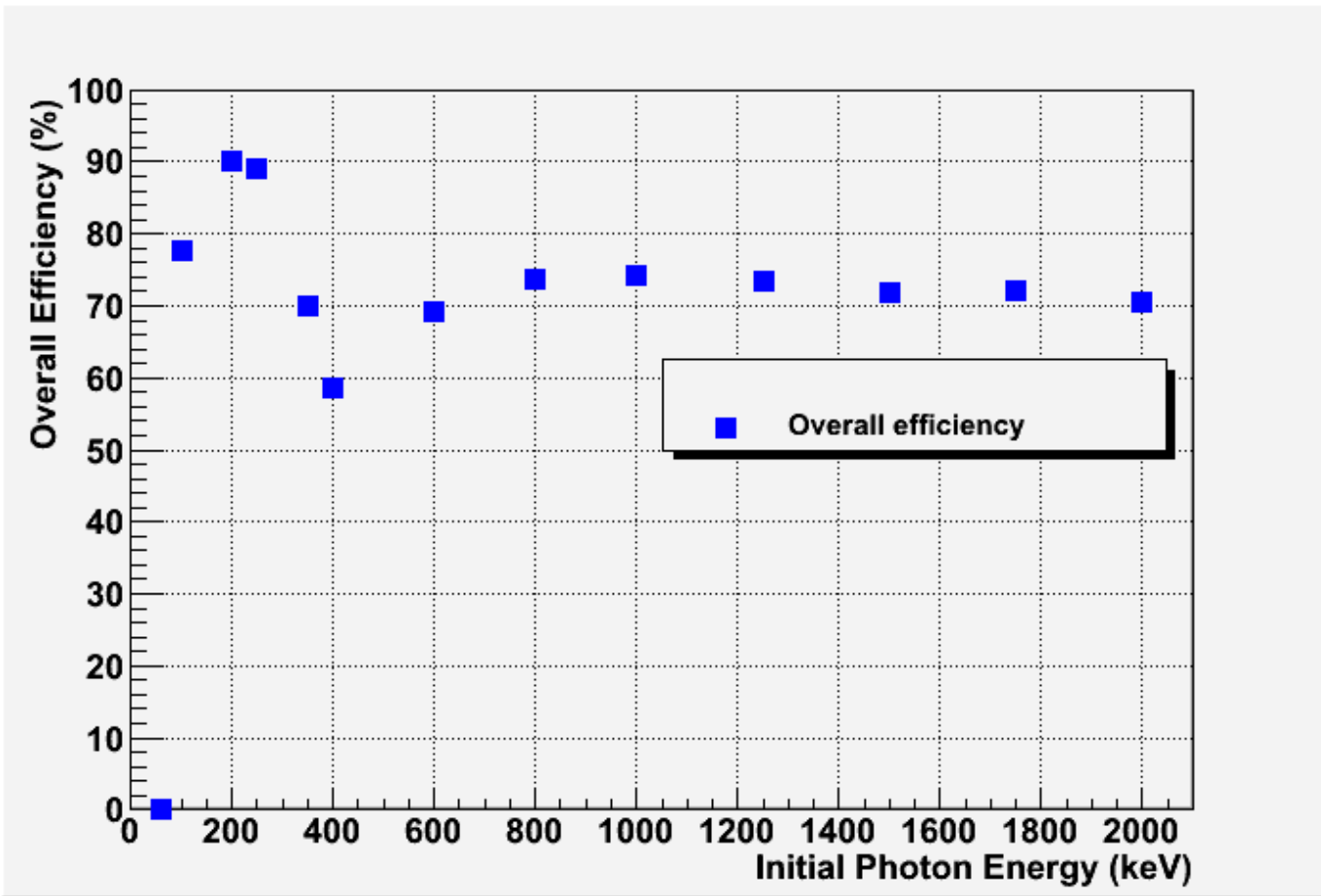

**FIGURE 17** Overall Compton sequence reconstruction efficiency using the DCS-B and MCS without constraints algorithms, as a function of the initial photon energy.

Although the above figure indicates that the sequence reconstruction efficiency is very high at low energies (below 300 keV), we do not expect this effect to be mirrored in the event reconstruction efficiency as well, since in this area of the spectrum the Compton kinematics are masked by the Doppler broadening effect.

## ACKNOWLEDGMENTS

This work is supported by the European Community's Seventh Framework Program FP7-SEC-2007-01.